\documentclass[12pt]{article}
\usepackage{epsfig}
\def\refb#1{(\ref{#1})}
\newcommand{\be}{\begin{equation}}
\newcommand{\ee}{\end{equation}} 
\newcommand{\ba}{\begin{eqnarray}}
\newcommand{\ea}{\end{eqnarray}}

\def\laq{\raise 0.4ex\hbox{$<$}\kern -0.8em\lower 0.62ex\hbox{$\sim$}}
\def\gaq{\raise 0.4ex\hbox{$>$}\kern -0.7em\lower 0.62ex\hbox{$\sim$}}
\begin{document}
\title{Brane factories}

\author{Eun-Joo Ahn$^{1}$\footnote{Email: sein@oddjob.uchicago.edu}, Marco
Cavagli\`a$^{2,3}$\footnote{Email: marco.cavaglia@port.ac.uk}, Angela V.\
Olinto$^{1,4}$\footnote{Email: olinto@oddjob.uchicago.edu}\\ \\
$^1$ \small Department of Astronomy and Astrophysics and Enrico Fermi
\small Institute, University of Chicago,\\
\small 5640 S.\ Ellis Avenue, Chicago, IL 60637 USA\\
$^{2}$ \small Center for Theoretical Physics,
\small Massachusetts Institute of Technology,\\
\small77 Massachusetts Avenue, Cambridge MA
02139-4307 USA \\
$^{3}$ \small Institute of Cosmology and Gravitation, University of
\small Portsmouth,\\ 
\small Portsmouth PO1 2EG, U.K.\\
$^{4}$ \small Center for Cosmological Physics, University of Chicago, Chicago,
IL 60637, USA\\}
\date{(MIT-CTP-3218)}
\maketitle
\begin{abstract}
We propose that higher-dimensional extended objects ($p$-branes) are created by
super-Planckian scattering processes in theories with TeV scale gravity. As an
example, we compute the cross section for $p$-brane creation in a
$(n+4)$-dimensional spacetime with asymmetric compactification. We find that
the cross section for the formation of a brane which is wounded on a compact
submanifold of size of the fundamental gravitational scale is larger than the
cross section for the creation of a spherically symmetric black hole.
Therefore, we predict that branes are more likely to be created than black holes
in super-Planckian scattering processes in these manifolds. The higher rate of
$p$-brane production has important phenomenological consequences, as it
significantly enhances possible detection of non-perturbative gravitational
events in future hadron colliders and cosmic rays detectors.
\end{abstract}
The fundamental Planck scale may be of the order TeV as in some models of extra
dimensions \cite{Antoniadis:1990ew,Arkani-Hamed:1998rs,Antoniadis:1998ig,
Randall:1999ee,Randall:1999vf,Antoniadis:2001sw,Benakli:1999yc}. In these
theories, processes at energies $\gaq$ TeV may experimentally test quantum
gravitational effects. In a series of recent papers 
\cite{Argyres:1998qn,Banks:1999gd,Giddings:2001bu,Dimopoulos:2001hw,
Giddings:2001ih,Feng:2001ib} it has been proposed that particle collisions with
center-of-mass energy larger than a few TeV and sufficiently small impact
parameter might generate black holes. The formation of super-Planckian black
holes and their subsequent evaporation would be detectable in future hadron
colliders \cite{Giddings:2001bu,Dimopoulos:2001hw,Giddings:2001ih} and in high
energy cosmic ray detectors as black holes would form in the Earth's atmosphere
\cite{Feng:2001ib,Anchordoqui:2001cg,Uehara:2001yk,Anchordoqui:2001ei,Ringwald:2002vk}.
\footnote{See Refs.\ \cite{Voloshin:2001fe,Voloshin:2001vs} for criticisms and
Refs.\ \cite{Rizzo:2001dk,Solodukhin:2002ui, Eardley:2002re} for counter
criticisms.} The detection of black hole formation can open a era in both 
experimental and theoretical high-energy physics. Thus the recent explosion of 
papers on the phenomenological aspects of black hole production 
is of no surprise. If super-Planckian scattering
probes non-perturbative quantum gravity, then formation of spherically
symmetric black holes is just the simplest of a plethora of possible
high-energy physical processes. At super-Planckian energies, we expect the
creation of {\it any} non-perturbative gravitational object which is predicted
by a given theory of quantum gravity. In particular, in the presence of extra
dimensions, one should expect the creation of higher-dimensional objects
($p$-branes) \cite{Stelle:1996tz,Stelle:nv}. Thus far, this exciting
possibility has been overlooked in the literature. 

In this paper we propose that $p$-branes are created by super-Planckian
scattering processes. To make our claim quantitative, we compute the cross
section for $p$-brane creation in a simple model. We assume a flat asymmetric
compactification for the extra dimensions. Asymmetric compactifications are 
suggested by some string theory models \cite{Lykken:1999ms,Antoniadis:1999rm}.
Specifically, we consider $m$ flat compact extra dimensions with size of order
of the fundamental scale $L_\star=M_{\star}^{-1}$ and $n-m$ flat extra
dimensions with size of order $L'\gg M_{\star}^{-1}$. (The generalization to
more than two different compactification scales is trivial.) Setting $n=m$ we
obtain the standard flat symmetric compactification of
\cite{Arkani-Hamed:1998rs,Antoniadis:1998ig}. We find that the cross section
for the formation of a $p$-brane whose dimensions are wound (``wrapped'') around
the $m$ extra dimensions is larger than the cross section for the formation of
spherically symmetric black holes. In this case, if super-Planckian scattering
processes lead to non-perturbative formation of gravitational objects, the rate
of formation of higher-dimensional branes is higher than the rate of formation
of spherically symmetric black holes. 

We consider for simplicity uncharged, non-spinning $p$-brane solutions of
$(n+4)$-dimensional Einstein gravity. The generalization of the model to either
the charged case or to string theory is straightforward and leads to no
significantly different results. Using standard notations, we write the
$(n+4)$-dimensional Einstein-Hilbert action
\be
S_{EH} = {M_{\star}^{n+2}\over 16 \pi}\int d^{n+4}x \sqrt{-g}\,{\cal R}(g)\,.
\label{EHaction}
\ee
The fundamental Planck scale $M_{\star}$ is related to the observed Planck
scale $M_{obs}\approx 10^{16}$ TeV by the relation
\be
M_{\star}=M_{obs}V_n^{-1/2}\,,
\label{Planck}
\ee
where $V_n$ is the volume of the extra dimensions in fundamental Planck units.
If $V_n\approx 10^{32}$, $M_{\star}$ is of order TeV.

An uncharged, static $p$-brane living in a $(n+4)$-dimensional spacetime
is described by the ansatz \cite{Gregory:1995qh}
\be
ds^2=A(r)(-dt^2+dz_i^2)+B(r)dr^2+C(r)d\Omega^2_{q}\,,
\label{pbrane}
\ee
where $z_i$ ($i=1,\dots,p$) are the brane coordinates and $d\Omega^2_{q}$
($q=n-p+2$) is the line element of the $q$-dimensional unit sphere.

The general solution of Eq.\ \refb{EHaction} with ansatz \refb{pbrane} has been
found in Ref.\ \cite{Gregory:1995qh} and later generalized to non
boost-symmetric configurations in Ref.\ \cite{Cavaglia:1997hc}. The metric is
\be
ds^2=R^{{\Delta\over p+1}}(-dt^2+dz_i^2)+R^{{2-q-\Delta\over
q-1}}dr^2+r^2R^{{1-\Delta\over q-1}}d\Omega^2_{q}\,,
\label{boosted}
\ee
where
\be
R(r)=1-\left({r_{p}\over r}\right)^{q-1}\,.
\label{R}
\ee
$\Delta$ is a constant parameter related to the brane dimension $p$
and to the sphere dimension $q$ by
\be
\Delta=\sqrt{{q(p+1)\over p+q}}\,.
\label{Delta}
\ee
The spherically symmetric solution is recovered for $p=0$. In this case
$\Delta=1$, and Eq.\ \refb{boosted} reduces to the $(n+4)$-dimensional
Schwarzschild black hole.

Let us briefly discuss two interesting features of Eq.\ \refb{boosted}. When
$p=0$ (black hole case) $r=r_{p}$ defines the Schwarzschild horizon. For
$p\not=0$ ($p$-brane case) the metric \refb{boosted} possesses a naked
singularity at $r=r_{p}$ which is the higher-dimensional analogue of a cosmic
string conical singularity \cite{Gregory:1995qh}. $r_{p}$ sets the curvature
scale of the geometry. Therefore $r_p$ can be interpreted as the physical
radius of the brane though the proper area of the $p$-brane per unit
brane-volume $V_p$ is infinite. The interpretation of the curvature singularity
has been discussed in Ref.\ \cite{Gregory:1995qh}. The metric \refb{boosted} is
interpreted as {\em vacuum} exterior solution to the $p$-brane, with the
curvature singularity being smoothed out by the core of the $p$-brane. In
analogy to the black hole case, we expect that a scattering process with impact
parameter $b\,\laq\, r_{p}$ will produce a $p$-brane which is described by a
suitable localized energy field configuration. Being an extended object endowed
with tension (mass/unit $p$-volume), the $p$-brane \refb{boosted} is unstable
\cite{Zwiebach}. The decay process of the $p$-brane depends on the type of
instability and is presently very speculative. String field theory arguments
\cite{Sen:1999mh,Sen:1999mg,Sen:1999xm,Moriyama:2000dc,Rastelli:2000hv,
Lee:2001cs,Lee:2001ey} suggest that the $p$-brane decays into lower dimensional
branes, and eventually into gauge radiation. On the other hand, analogy to
cosmic strings \cite{Eardley:1995au,Hawking:1995zn,Gregory:1995hd} suggets a
nonperturbative instability, which would unlikely lead to a final fragmentation
into $0$-branes. For the bosonic uncharged $p$-brane \refb{boosted}, the main
difference from the black hole scenario is that the absence of an event horizon
does not immediately lead to Hawking evaporation, though we expect the
$p$-branes to eventually evaporate by emission of observable, possibly thermal,
particles. The intermediate states of the $p$-brane decay are highly dependent
on the details of the theory considered. However we do not expect significant
qualitative differences as far as the final evaporation stage is concerned. For
instance, when the model is embedded in string theory, the presence of charges
usually lead to $p$-branes with horizons. As long as the solution does not
saturate the Bogomol'ny bound \cite{Stelle:1996tz,Stelle:nv} the $p$-brane will
have nonzero entropy and will evaporate by Hawking radiation.

$p$-branes form when two partons $i,j$ with center-of-mass energy
$E_{ij}=\sqrt{s}$ scatter with impact parameter $b\,\laq\, r_{p}$. The
geometrical cross section for this process can be approximated by an absorptive
black disk with area $\pi r_{p}^2$, i.e.,
\be
\sigma_{ij\to br}(s;n,p)=F(s)\pi r_{p}^2\,,
\label{sigma}
\ee
where $F(s)$ is a dimensionless form factor of order one. By analogy to the
black hole case, in the following we will assume $F(s)=1$. This is a rather
conservative choice that has been widely discussed in the literature (see
Refs.\ \cite{Anchordoqui:2001cg,Solodukhin:2002ui,Eardley:2002re}). From Eq.\
\refb{boosted} the radius of a $p$-brane with mass $M_{p}$ is
\be
\displaystyle
r_{p} = {1\over\sqrt{\pi}M_{\star}} \gamma(n,p)\,V_p^{-{w\over n+1}}
\left({M_{p}\over M_{\star}}\right)^{w\over n+1}\,.
\label{rp}
\ee
$V_p$ is the volume of the extra dimensions in fundamental Planck units
where the $p$-brane wraps, $w=[1-p/(n+1)]^{-1}$, and
\be
\displaystyle
\gamma(n,p)=\left[{8\,\Gamma \left(\displaystyle{n+3-p\over 2}\right)\over
(2+n)\sqrt{(p+1)^{-1}\left(1-\displaystyle{p\over n+2}\right)}}\right]^{w\over
n+1} \,.
\label{gamma}
\ee
The cross section for $p$-brane formation is
\be
\displaystyle
\sigma_{ij\to br}(s;p,n,V_p) \approx {1\over s_{\star}}\gamma(n,p)^2\,
 V_p^{-{2w\over n+1}} \left({s\over s_{\star}}\right)^{w\over n+1}\,,
\label{cross-brane}
\ee
where $s_{\star}=M_{\star}^2$. Let us compare Eq.\ \refb{cross-brane} to the
cross section for the production of a spherically symmetric black hole with
mass $M_p$ (see e.g.\ Ref.\ \cite{Feng:2001ib}). The latter is recovered for
$p=0$ and is explicitly given by
\be
\displaystyle
\sigma_{ij\to bh}(s;n)\approx {1\over s_{\star}}
\gamma(n,0)^2\left({s\over s_{\star}}\right)^{1\over n+1}\,.
\label{cross-bh}
\ee
The ratio of the two cross sections is
\be\label{ratio1}
\Sigma(s;n,p,V_p)\equiv{\sigma_{ij\to br}\over\sigma_{ij\to bh}}\approx
V_p^{-{2w\over n+1}}{\gamma(n,p)^2\over\gamma(n,0)^2}
\left({s\over s_{\star}}\right)^{{w-1\over n+1}}\,.
\ee
Since $w>1$ for any $n\,\ge p>0$, $\Sigma$ becomes larger for higher energy. At
fixed $s$, the value of $\Sigma$ depends on the dimensionality of the brane and
on the size of the extra dimensions. In the scenario with $m$ extra dimensions
compactified on the $L$ scale and $n-m$ dimensions compactified on the $L'$
scale, Eq.\ \refb{Planck} gives
\be
\left({L\over L_{\star}}\right)^{m}
\left({L'\over L_{\star}}\right)^{n-m}=
\left({M_{obs}\over M_{\star}}\right)^{2}\,.
\ee
If we assume that the $p$-brane wraps on $r$ small dimensions ($r\le m)$
and on $p-r$ large dimensions, the volume $V_p$ is
\be\label{Vp}
V_p=\left({L\over L_{\star}}\right)^{r}
\left({L'\over L_{\star}}\right)^{p-r}=\left({L\over
L_{\star}}\right)^{nr-mp\over
n-m}\left({M_{obs}\over M_{\star}}\right)^{2(p-r)\over n-m}\,.
\ee
Substituting Eq.\ \refb{Vp} in Eq.\ \refb{ratio1} we find
\be\label{ratio2}
\Sigma(s;n,m,p,r)\approx
\left({M_{obs}\over M_{\star}}\right)^{-\alpha}
\left({L\over L_{\star}}\right)^{-\beta}{\gamma(n,p)^2\over\gamma(n,0)^2}
\left({s\over s_{\star}}\right)^{{w-1\over n+1}}\,,
\ee
where
\be\label{alphabeta}
\alpha={4(p-r)\over (n-m)(n-p+1)}\ge 0\,,\qquad
\beta={2(nr-mp)\over (n-m)(n-p+1)}\ge 0\,.
\ee
In theories with TeV scale gravity, $M_{obs}/M_{\star}\approx 10^{14}$
($10^{16}$) for $M_{\star}\approx 100$ TeV ($1$ TeV). Since $0\le
(w-1)/(n+1)\le 1$, for physically interesting energy scales the $p$-brane cross
section is suppressed w.r.t.\ spherically symmetric black hole cross section by
a factor $\approx 10^{14\alpha}$ ($10^{16\alpha}$). The largest cross section
is obtained for $p=r$, i.e., when the $p$-brane is completely wrapped on
small-size dimensions:
\be\label{ratio3}
\Sigma(s;n,m,p\le m)\approx
\left({L\over L_{\star}}\right)^{-{2p\over
n-p+1}}{\gamma(n,p)^2\over\gamma(n,0)^2}
\left({s\over s_{\star}}\right)^{{w-1\over n+1}}\,,
\ee
Since $L\laq\, L_{\star}$, the $p$-brane formation process dominates the black
hole formation process. When the $p$-brane is wrapped on some of the large
extra dimensions, the $p$-brane cross section is instead suppressed w.r.t.\
black hole cross section. $\Sigma$ slightly increases with the dimension of the
brane. Therefore, in a spacetime with $m$ fundamental-scale extra dimensions
and $n-m$ large extra dimensions a $m$-brane is the most likely object to be
created.

In the following, we give a 11-dimensional spacetime as a concrete example.
Let us consider $m=5$ fundamental-scale extra
dimensions $L=L_{\star}=10^{-2}~({\rm TeV})^{-1}$ and two large extra
dimensions $L\approx 10^{12}~({\rm TeV})^{-1}$. At $s\approx 10 s_\star$ the
cross sections for the formation of a 5-brane and a 4-brane completely wrapped
on the fundamental-size dimensions are enhanced by a factor $\approx 2$ and
$\approx 1.5$ w.r.t.\ cross section for creation of a spherically symmetric
black hole, respectively. If the 5-brane wraps on four extra dimensions with
fundamental scale size and on one large extra dimension, $\Sigma(s\approx 10
s_\star)$ is suppressed by a factor $\approx 10^{8}$.
\begin{center}
\null
\vskip -0.4truein
\hskip -0.4truein
\epsfig{file=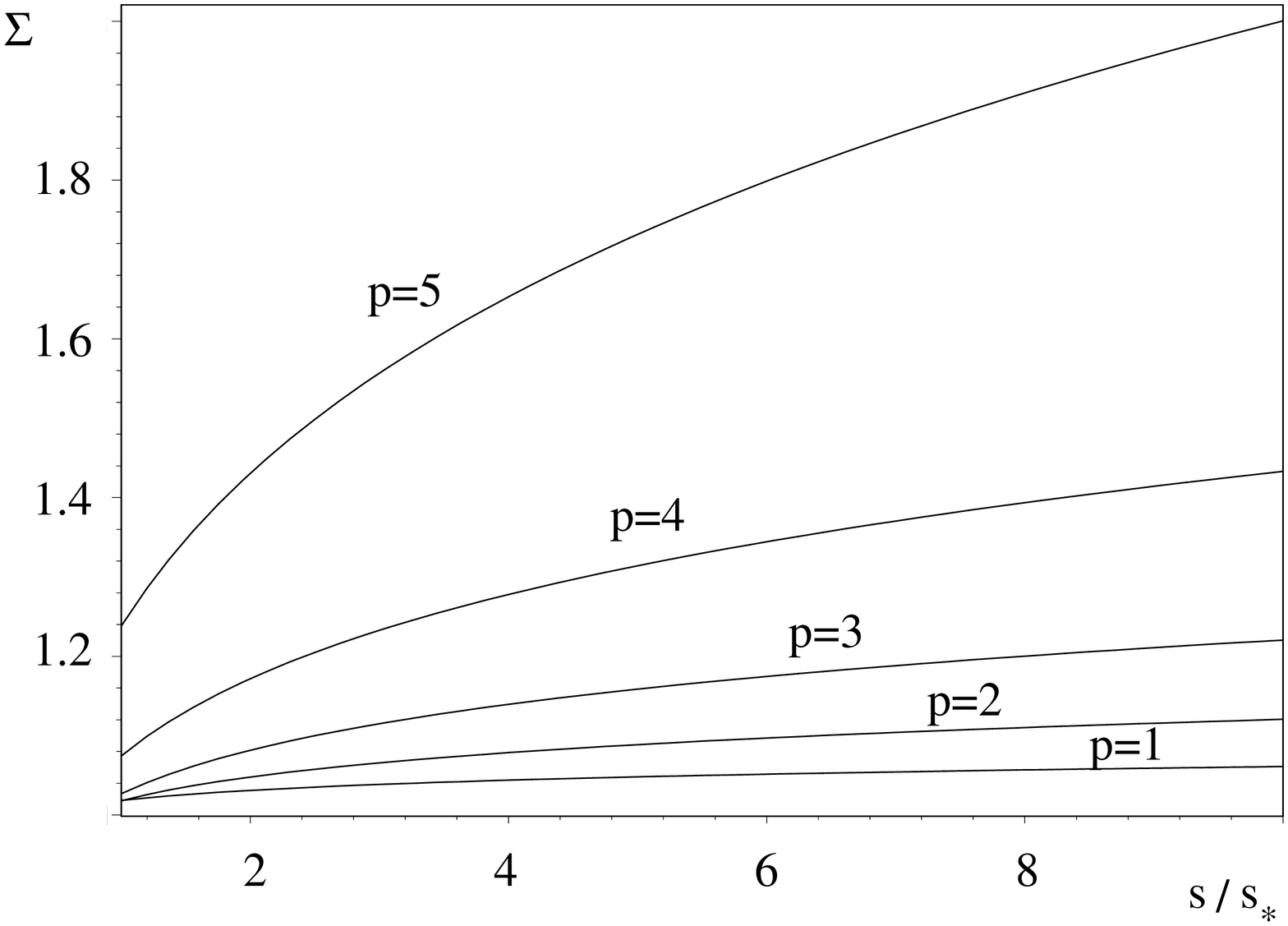,angle=270,width=5truein}\\
\end{center}
\vskip -0.4truein
{\small
Fig.\ 1: Ratio between the cross section for the creation of $p$-branes
($p\,\le\, m$) completely wrapped on fundamental-size dimensions and a
spherically symmetric black hole in a spacetime with $m=5$ fundamental-size
extra
dimensions $L=L_{\star}=10^{-2}~({\rm TeV})^{-1}$ and $n-m=2$ large extra
dimensions of size $L'\approx 10^{12}~({\rm TeV})^{-1}>>L_{\star}$.}
\vskip 0.1truein
The cross sections are enhanced if the dimensions where the $p$-brane is
wrapped are smaller than the fundamental scale. For instance, assuming  $L=0.5
L_{\star}$ ($L=0.25 L_{\star}$) the cross section for the creation of
$5$-branes in a $11$-dimensional spacetime is enhanced by a factor $\approx 10$
(100). This result is of special interest to the ultra high energy
cosmic ray community, as the
enhancement of the cross section should allow a sufficient flux of $p$-branes
to be detected by ground array and air fluorescence detectors \cite{Nagano:ve}.
\begin{center}
\null
\vskip -0.4truein
\hskip -0.4truein
\epsfig{file=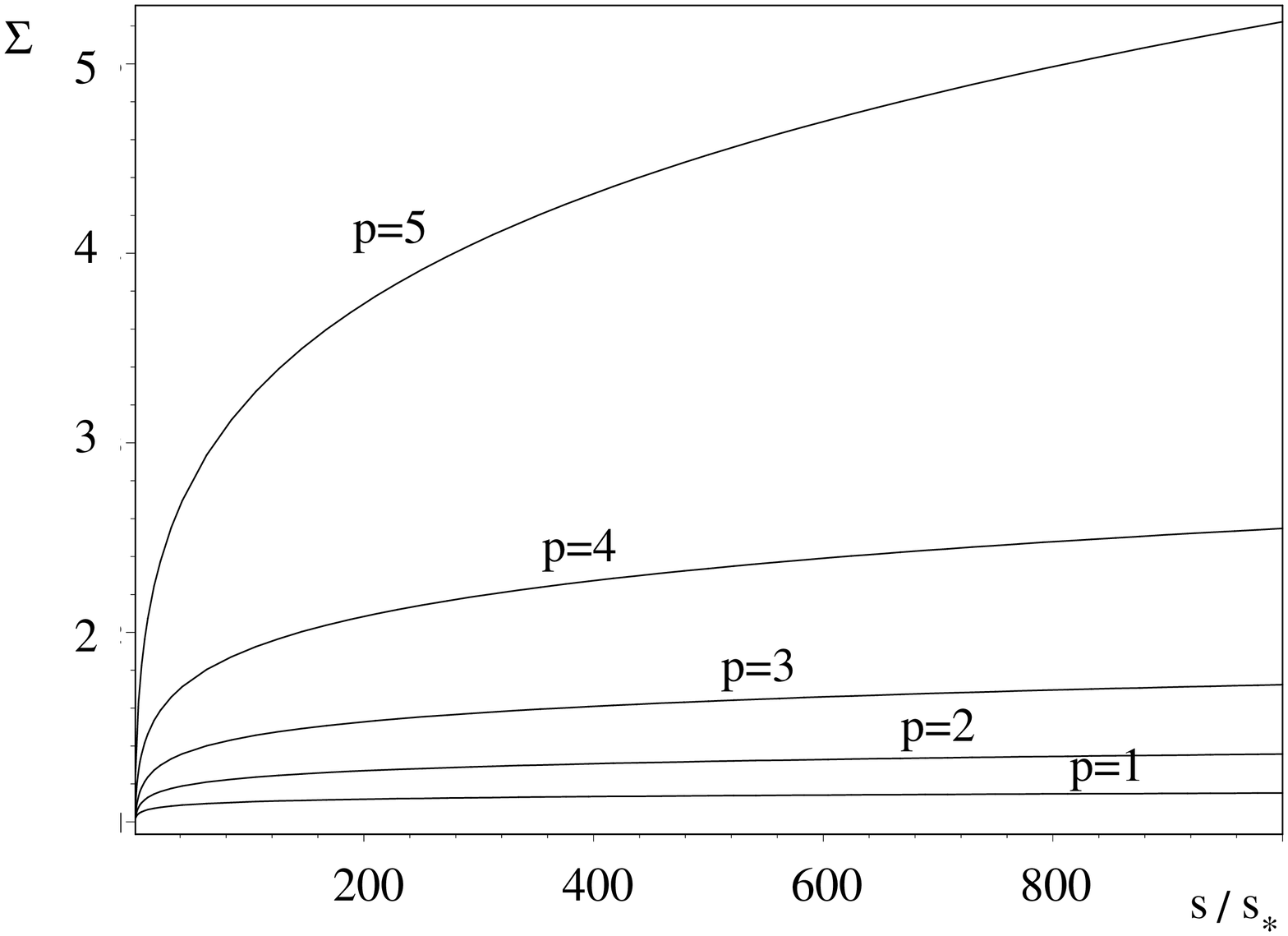,angle=270,width=5truein}\\
\end{center}
\vskip -0.4truein
{\small
Fig.\ 2: Ratio between the cross section for the creation of $p$-branes
($p\,\le\, m$) completely wrapped on fundamental-size dimensions and a
spherically symmetric black hole in a spacetime with $m=5$ fundamental-size
extra
dimensions $L=L_{\star}=10^{-1}~({\rm TeV})^{-1}$ and $n-m=2$ large extra
dimensions of size $L'\approx 10^{14}~({\rm TeV})^{-1}\approx 2\cdot
10^{-3}~{\rm cm}$. If the fundamental-size extra dimensions have size $L=0.25
L_{\star}$ the cross sections are enhanced by a factor $\approx 100$, 16, 5, 2,
1.5 for $p=5$, 4, 3, 2 and 1, respectively.}
\vskip 0.1truein
Finally, let us consider the case where all extra dimensions are compactified
on the same scale $L'$. We have
\be\label{ratio4}
\Sigma\approx
\left({M_{obs}\over M_{\star}}\right)^{-{4wp\over
n(n+1)}}{\gamma(n,p)^2\over\gamma(n,0)^2}
\left({s\over s_{\star}}\right)^{{w-1\over n+1}}\,.
\ee
In this case the cross section for $p$-brane formation is subdominant to the
cross section for black hole formation. This result is understood
qualitatively as follows. If all the extra dimensions have (large) identical
characteristic size, the spacetime appears isotropic to the $p$-brane and a
spherically symmetric object is likely to form. Conversely, when the
compactification is asymmetric, that is, $m$ of the extra dimensions are
smaller than the others, non-spherically symmetric objects are more likely to
be created. The most likely $p$-brane to form is that with the highest symmetry
compatible with spacetime symmetries, i.e., a $m$-brane.

To conclude, let us briefly comment on the relevance of our results for
short-distance experimental physics. In Ref.\ \cite{Giddings:2001bu} it has
been argued that the creation of event horizons by relativistic high-energy
collisions limits the ability to probe short-distance physics by perturbative
hard scattering processes in future colliders. Since neutral $p$-branes do not
possess an event horizon, $p$-brane formation does not cloak hard processes.
Therefore, different hard super-Planckian processes can still lead to different
experimental signatures depending on the physics of the collision and on the
structure of the extra dimensions. Rather than representing the {\em end of
experimental investigation of short-distance physics} \cite{Giddings:2001bu},
detection of $p$-branes may represent the {\em beginning of experimental
quantum gravity}.
\section*{Acknowledgements}
We are very grateful to Ignatios Antoniadis, Gia Dvali, Daniele Fargion, 
Jonathan Feng, Haim Goldberg, Ruth Gregory, Amihay Hanany, Joseph Polchinski, 
Craig Tyler, Alex Vilenkin and Barton Zwiebach, for interesting discussions 
and useful comments. E.-J.~A.\ and M.~C.\
thank the Center for Theoretical Physics of the Massachusetts Institute of
Technology and the Department of Astronomy and Astrophysics of the University
of Chicago for the kind hospitality, respectively. This work is supported in
part by funds provided by the U.S.\ Department of Energy under cooperative
research agreement DE-FC02-94ER40818 at MIT and grant DE-FG0291-ER40606 at the
University of Chicago and by the National Science Foundation grant NSF
PHY-0114422 at the Center for Cosmological Physics at the University of
Chicago.


\thebibliography{99}
%
\bibitem{Antoniadis:1990ew}
I.~Antoniadis,
Phys.\ Lett.\ B {\bf 246}, 377 (1990).

\bibitem{Arkani-Hamed:1998rs}
N.~Arkani-Hamed, S.~Dimopoulos and G.~R.~Dvali,
Phys.\ Lett.\ B {\bf 429}, 263 (1998)
[arXiv:hep-ph/9803315].

\bibitem{Antoniadis:1998ig}
I.~Antoniadis, N.~Arkani-Hamed, S.~Dimopoulos and G.~R.~Dvali,
Phys.\ Lett.\ B {\bf 436}, 257 (1998)
[arXiv:hep-ph/9804398].

\bibitem{Randall:1999ee}
L.~Randall and R.~Sundrum,
Phys.\ Rev.\ Lett.\  {\bf 83}, 3370 (1999)
[arXiv:hep-ph/9905221].

\bibitem{Randall:1999vf}
L.~Randall and R.~Sundrum,
Phys.\ Rev.\ Lett.\  {\bf 83}, 4690 (1999)
[arXiv:hep-th/9906064].

\bibitem{Antoniadis:2001sw}
I.~Antoniadis, S.~Dimopoulos and A.~Giveon,
JHEP {\bf 0105}, 055 (2001)
[arXiv:hep-th/0103033].

\bibitem{Benakli:1999yc}
K.~Benakli and Y.~Oz,
Phys.\ Lett.\ B {\bf 472}, 83 (2000)
[arXiv:hep-th/9910090].

\bibitem{Argyres:1998qn}
P.~C.~Argyres, S.~Dimopoulos and J.~March-Russell,
Phys.\ Lett.\ B {\bf 441}, 96 (1998)
[arXiv:hep-th/9808138].

\bibitem{Banks:1999gd}
T.~Banks and W.~Fischler,
arXiv:hep-th/9906038.

\bibitem{Giddings:2001bu}
S.~B.~Giddings and S.~Thomas,
Phys.\ Rev.\ D {\bf 65}, 056010 (2002)
[arXiv:hep-ph/0106219].

\bibitem{Dimopoulos:2001hw}
S.~Dimopoulos and G.~Landsberg,
Phys.\ Rev.\ Lett.\  {\bf 87}, 161602 (2001)
[arXiv:hep-ph/0106295].


\bibitem{Giddings:2001ih}
S.~B.~Giddings,
in {\it Proc. of the APS/DPF/DPB Summer Study on the Future of Particle Physics (Snowmass 2001) } ed. N.~Graf,
arXiv:hep-ph/0110127.

\bibitem{Feng:2001ib}
J.~L.~Feng and A.~D.~Shapere,
Phys.\ Rev.\ Lett.\  {\bf 88}, 021303 (2002)
[arXiv:hep-ph/0109106].

\bibitem{Anchordoqui:2001cg}
L.~A.~Anchordoqui, J.~L.~Feng, H.~Goldberg and A.~D.~Shapere,
Phys.\ Rev.\ D {\bf 65}, 124027 (2002)
[arXiv:hep-ph/0112247].

\bibitem{Uehara:2001yk}
Y.~Uehara,
Prog.\ Theor.\ Phys.\  {\bf 107}, 621 (2002)
[arXiv:hep-ph/0110382].

\bibitem{Anchordoqui:2001ei}
L.~Anchordoqui and H.~Goldberg,
Phys.\ Rev.\ D {\bf 65}, 047502 (2002)
[arXiv:hep-ph/0109242].

\bibitem{Ringwald:2002vk}
A.~Ringwald and H.~Tu,
Phys.\ Lett.\ B {\bf 525}, 135 (2002)
[arXiv:hep-ph/0111042].

\bibitem{Voloshin:2001fe}
M.~B.~Voloshin,
Phys.\ Lett.\ B {\bf 524}, 376 (2002)
[arXiv:hep-ph/0111099].

\bibitem{Voloshin:2001vs}
M.~B.~Voloshin,
Phys.\ Lett.\ B {\bf 518}, 137 (2001)
[arXiv:hep-ph/0107119].


\bibitem{Rizzo:2001dk}
T.~G.~Rizzo,
in {\it Proc. of the APS/DPF/DPB Summer Study on the Future of Particle Physics (Snowmass 2001) } ed. N.~Graf,
arXiv:hep-ph/0111230.

\bibitem{Solodukhin:2002ui}
S.~N.~Solodukhin,
Phys.\ Lett.\ B {\bf 533}, 153 (2002)
[arXiv:hep-ph/0201248].

\bibitem{Eardley:2002re}
D.~M.~Eardley and S.~B.~Giddings,
Phys.\ Rev.\ D {\bf 66}, 044011 (2002)
[arXiv:gr-qc/0201034].

\bibitem{Stelle:1996tz}
K.~S.~Stelle,
arXiv:hep-th/9701088.

\bibitem{Stelle:nv}
K.~S.~Stelle,
{\it Given at APCTP Winter School on Dualities of Gauge and String
Theories, Seoul and Sokcho, Korea, 17-28 Feb 1997}.

\bibitem{Lykken:1999ms}
J.~Lykken and S.~Nandi,
Phys.\ Lett.\ B {\bf 485}, 224 (2000)
[arXiv:hep-ph/9908505].

\bibitem{Antoniadis:1999rm}
I.~Antoniadis and B.~Pioline,
Nucl.\ Phys.\ B {\bf 550}, 41 (1999)
[arXiv:hep-th/9902055].

\bibitem{Gregory:1995qh}
R.~Gregory,
Nucl.\ Phys.\ B {\bf 467}, 159 (1996)
[arXiv:hep-th/9510202].

\bibitem{Cavaglia:1997hc}
M.~Cavagli\`a,
Phys.\ Lett.\ {\bf B413}, 287 (1997)
[hep-th/9709055].

\bibitem{Zwiebach}
We are grateful to B.~Zwiebach for this remark.

\bibitem{Sen:1999mh}
A.~Sen,
Int.\ J.\ Mod.\ Phys.\ A {\bf 14}, 4061 (1999)
[arXiv:hep-th/9902105].

\bibitem{Sen:1999mg}
A.~Sen,
arXiv:hep-th/9904207.

\bibitem{Sen:1999xm}
A.~Sen,
JHEP {\bf 9912}, 027 (1999)
[arXiv:hep-th/9911116].

\bibitem{Moriyama:2000dc}
S.~Moriyama and S.~Nakamura,
Phys.\ Lett.\ B {\bf 506}, 161 (2001)
[arXiv:hep-th/0011002].

\bibitem{Rastelli:2000hv}
L.~Rastelli, A.~Sen and B.~Zwiebach,
Adv.\ Theor.\ Math.\ Phys.\  {\bf 5}, 353 (2002)
[arXiv:hep-th/0012251].

\bibitem{Lee:2001cs}
T.~Lee,
Phys.\ Lett.\ B {\bf 520}, 385 (2001)
[arXiv:hep-th/0105264].

\bibitem{Lee:2001ey}
T.~Lee,
Phys.\ Rev.\ D {\bf 64}, 106004 (2001)
[arXiv:hep-th/0105115].

\bibitem{Eardley:1995au}
D.~M.~Eardley, G.~T.~Horowitz, D.~A.~Kastor and J.~Traschen,
Phys.\ Rev.\ Lett.\  {\bf 75}, 3390 (1995)
[arXiv:gr-qc/9506041].

\bibitem{Hawking:1995zn}
S.~W.~Hawking and S.~F.~Ross,
Phys.\ Rev.\ Lett.\  {\bf 75}, 3382 (1995)
[arXiv:gr-qc/9506020].

\bibitem{Gregory:1995hd}
R.~Gregory and M.~Hindmarsh,
Phys.\ Rev.\ D {\bf 52}, 5598 (1995)
[arXiv:gr-qc/9506054].

\bibitem{Nagano:ve}
M.~Nagano and A.~A.~Watson,
Rev.\ Mod.\ Phys.\  {\bf 72}, 689 (2000).
\end{document}